\input{aipcheck}

\documentclass[
    ,final            
  ]
  {aipproc}

\layoutstyle{6x9}

\newbox\grsign \setbox\grsign=\hbox{$>$} \newdimen\grdimen \grdimen=\ht\grsign
\newbox\simlessbox \newbox\simgreatbox \newbox\simpropbox
\setbox\simlessbox=\hbox{\raise.5ex\hbox{$<$}\llap
     {\lower.5ex\hbox{$\sim$}}}\ht2=\grdimen\dp2=0pt
\def\simlt{\mathrel{\copy\simlessbox}}

\def\be{\begin{equation}}
\def\ee{\end{equation}}

\usepackage{epsfig}

\begin{document}

\title{Neutrino-cooled Accretion Disks around Spinning Black Holes}


\author{Wen-xin Chen}{
   address={Physics Department and Columbia Astrophysics Laboratory, 
            Columbia University, 538 West 120th Street, New York, NY 10027}
}

\author{Andrei M. Beloborodov}{
   address={Physics Department and Columbia Astrophysics Laboratory, 
            Columbia University, 538 West 120th Street, New York, NY 10027}
}

\begin{abstract}
We calculate the structure of accretion disk around a spinning black hole 
for accretion rates $\dot{M}=0.01 - 10 M_\odot$~$s^{-1}$. The model is fully 
relativistic and treats accurately the disk microphysics including neutrino 
emissivity, opacity, electron degeneracy, and nuclear composition. We find 
that the accretion flow always regulates itself to a mildly degenerate state 
with the proton-to-nucleon ratio $Y_e\simlt 0.1$ and becomes very neutron-rich. 
The disk has a well defined ``ignition'' radius where neutrino flux raises 
dramatically, cooling becomes efficient, and $Y_e$ suddenly drops. 
We also calculate other characteristic radii of the disk, including the
$\nu$-opaque and $\nu$-trapping radii, and show their dependence on $\dot{M}$.
Accretion disks around fast-rotating black holes produce intense neutrino 
fluxes which may deposit enough energy above the disk to generate a GRB jet. 
\end{abstract}

\maketitle



\small Most GRB models assume a short-lived accretion disk as a primary source
of energy. Such disks have huge accretion rates 
$\dot{M}= 0.01-10 M_\odot$~$s^{-1}$ and their 
cooling mechanism is neutrino emission \cite{Popham,Belo,Kohri}.
Special features of neutrino-cooled disks are: 

\medskip

--- The main cooling is due to reactions
$p+e^{-}\rightarrow n+\nu$ and $n+e^{+}\rightarrow p+\bar\nu$. \\
\indent --- Electron degeneracy affects the disk structure.  \\
\indent --- The disk can be opaque to neutrinos which affects the cooling 
rate.\\ 
\indent --- Nuclear composition changes with radius, consuming energy.  \\
\indent --- Neutrino cooling and electron degeneracy lead 
to very high neutron richness.

\medskip

\noindent
The disk model may be formulated following the classical vertically-integrated
approach of Shakura \& Sunyaev and approximating the azimuthal motion
in the disk, $u^\phi$, as Keplerian rotation. A small radial velocity
is superimposed on this rotation. It is approximately given by
$u^r=\alpha B(H/r)^2ru^\phi$ where $H$ is half-thickness of the 
disk (found from hydrostatic balance), $B$ is a numerical factor
determined by Kerr metric, and $\alpha=0.1-0.01$.

\medskip

The main equations of the accretion disk are as follows.
\small
\be 
  {\rm Energy~conservation:}
\hspace*{1cm}
   F^{+}-F^{-} = cu^r \left[ \frac{d(UH)}{dr}
        -\frac{U+P}{\rho}\frac{d(\rho H)}{dr} \right] . 
\hspace*{1.6cm}
\ee 
\small
\be 
 {\rm ~~~~Lepton~conservation:} 
\hspace*{0.8cm}
 \frac{1}{H}(\dot N_{\bar\nu}-\dot N_{\nu})
       =cu^r\left[ \frac{\rho}{m_p}\frac{dY_e}{dr}
     +\frac{d}{dr}(n_\nu -n_{\bar\nu}) \right] . 
\hspace*{1.5cm}
\ee
\small
\be 
 {\rm Charge~conservation:} 
\hspace*{1.cm}
  n_{e^-}-n_{e^+}=Y_e \frac{\rho}{m_p} . 
\hspace*{5.5cm}
\ee
\small
\be 
 {\rm Chemical~balance:} 
\hspace*{1cm}
  \mu_e-\mu_\nu =kT\ln\left( \frac{1-Y_e}{Y_e}\right) + (m_n-m_p)c^2 . 
\hspace*{1.6cm}
\ee
\rm
Here $F^{+}$ is the viscously dissipated energy per unit area, $F^{-}$ 
is the neutrino energy flux, $\dot N_{\nu}$ and $\dot N_{\bar\nu}$ are 
the number fluxes of neutrinos and antineutrinos,
$\rho$ is the mass density, $T$ is temperature, $Y_e$ is the
proton-to-baryon ratio, $\mu_e$ and $\mu_\nu$ are the chemical potentials 
of electrons and neutrinos, $n_{e^-}$ and $n_{e^+}$ are number densities
of electrons and positrons.
The energy density $U$ and pressure $P$ include contributions of 
baryons, electrons, positrons, radiation, and neutrinos. They are 
calculated by integrating the corresponding distribution functions
(Fermi-Dirac distributions for $e^\pm$ and $\nu,\bar{\nu}$ if the disk 
is opaque to neutrinos). 

Note that we include the advection terms in the laws of energy and 
lepton conservation (right-hand side of eqs.~1 and 2). They describe 
the most important effects of advection. Other effects of heat 
storage in the disk --- in particular, the dynamical effects of the 
radial pressure gradient on $u^r$ and $u^\phi$ --- are relatively small 
and neglected here, i.e. the standard $\alpha$-disk model is used to 
calculate $u^\phi$ and $u^r$.

The chemical balance and $\mu_\nu$ are used 
only in the $\nu$-opaque region; then the cooling rate is calculated 
as $F^-=cU_\nu/\tau_\nu+cU_{\bar{\nu}}/\tau_{\bar{\nu}}$. In the 
$\nu$-transparent region, $F^-$ equals 
$2H(\dot{U}_{{e^-}p}+\dot{U}_{{e^+}n})$ where $\dot{U}_{{e^-}p}$ and 
$\dot{U}_{{e^+}n}$ are the rates of energy release by reactions 
$e^-+p\rightarrow n+\nu$ and $e^++n\rightarrow p+\bar{\nu}$.
The abundance of $\alpha$-particles is determined by the nuclear
statistical equilibrium.

We have solved the set of disk equations numerically starting from an outer 
region where no neutrino cooling takes place and assuming that the infalling 
matter is initially made of $\alpha$ particles (heavier elements are 
decomposed into $\alpha$-particles as the matter approaches the black hole). 
A schematic picture of the accretion disk and its characteristic radii are
shown in Figure~1.

\begin{figure}[h] 
\epsfig{file=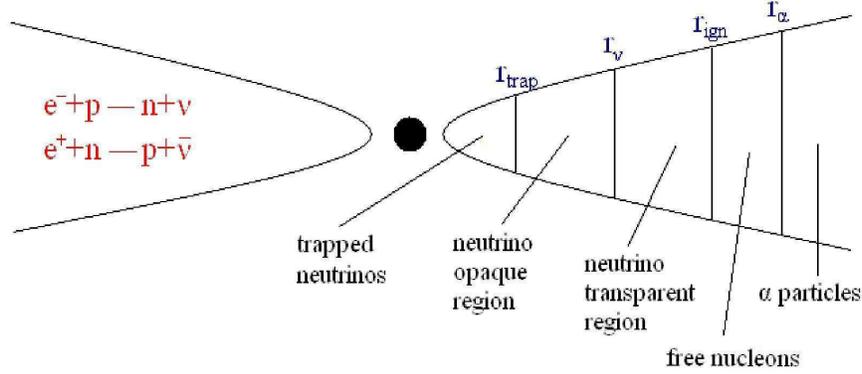, height=15pc, width=28pc} 
\caption{Schematic picture of the disk with characteristic radii indicated.}
\end{figure}

The outer region $r>50r_g$ ($r_g=2GM/c^2$ is the gravitational radius of 
the black hole) is dominated by $\alpha$ particles. Neutrino cooling is 
negligible in this region and the viscously dissipated heat is stored in 
the disk and advected inward. The destruction of $\alpha$ particles at 
$r_\alpha \sim 50 r_g$ consumes $\sim 7$~MeV per baryon, which makes the 
disk somewhat thinner. At the "ignition" radius $r_{ign}$ (slightly smaller 
than $r_\alpha$), $kT$ reaches $\sim$~MeV and neutrino cooling becomes 
significant, further reducing disk scale-height $H/r$. With decreasing 
radius the disk becomes opaque to neutrinos (first to $\nu$ and then to 
$\bar \nu$). In the innermost region of disks with $\dot{M}>2M_\odot$~s$^{-1}$
the neutrinos are trapped and advected toward the black hole.  
Figure~2 shows the found characteristic radii as functions of 
$\dot{M}$ for disks with $\alpha=0.1$ around 
black holes with spin parameters $a=0$ and $0.95$. 

\begin{figure}[h] 
\centering 
$\begin{array}{cc} \epsfig{file=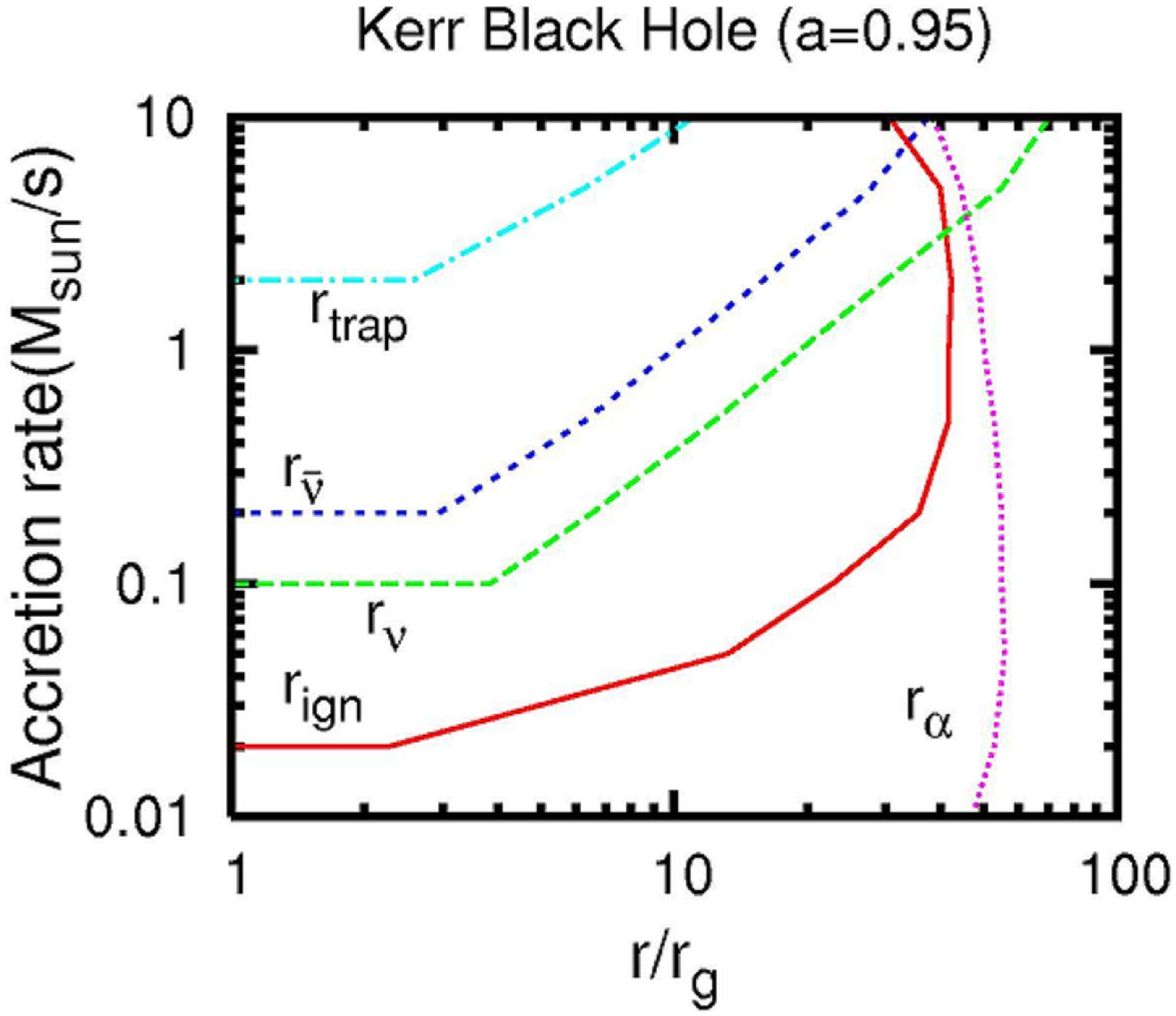, height=14pc, width=18pc} &
\epsfig{file=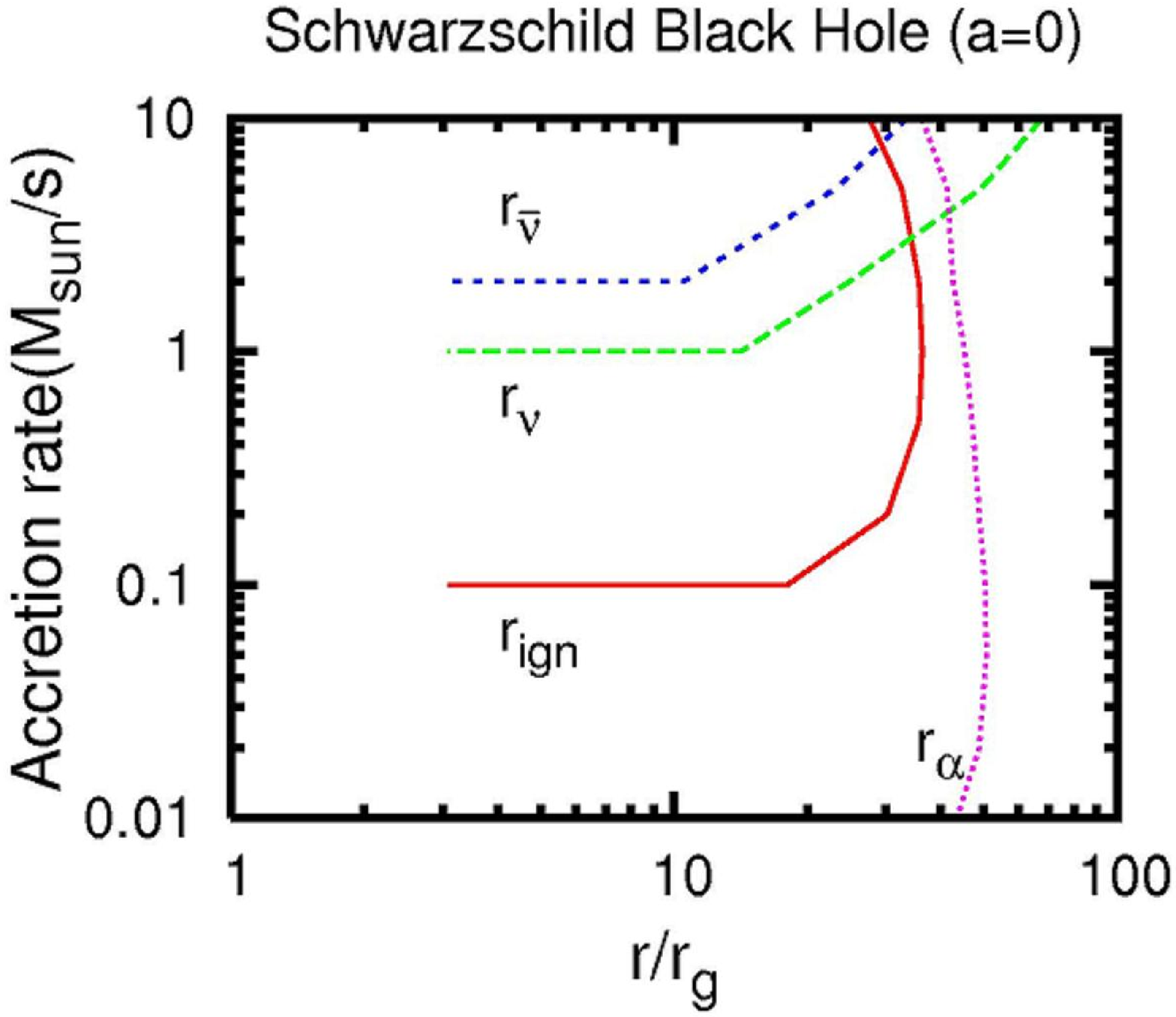, height=14pc, width=18pc} \end{array}$ 
\caption{
Characteristic radii in the accretion disk as a function of $\dot{M}$
for a black hole of mass $M=3M_\odot$ and spin parameter $a=0.95$ (left)
and 0 (right). Viscosity parameter $\alpha=0.1$ is assumed.
Neutrino cooling does not ignite when $\dot{M}$ is below
$0.02 M_\odot s^{-1}$ for Kerr and $0.1 M_\odot s^{-1}$ for 
Schwarzschild black holes.
Neutrino trapping takes place for Kerr black holes when 
$\dot{M}>2M_\odot$~s$^{-1}$.
The disk extends down to the marginally stable orbit, which is $3r_g$
for $a=0$ and $\approx r_g$ for $a=0.95$.
} 
\end{figure} 

We find two general properties of the disk:

\noindent 
1. --- It regulates itself to a mildly degenerate state with 
$\mu_e =1\sim 3 kT$. The reason of this regulation is the 
negative feedback of degeneracy and neutronization on the cooling rate: 
higher degeneracy $\mu_e /kT$ $\rightarrow$ fewer positrons 
($n_+/n_-\sim e^{-\mu_e/kT}$) $\rightarrow$ lower equilibrium rate of 
neutrino emission $\rightarrow$ lower cooling rate $\rightarrow$ higher 
temperature $\rightarrow$ lower degeneracy. 

\noindent 
2. --- In a broad range of relevant accretion rates the disk is found to
reach $Y_e \sim 0.1$. It corresponds to a very high neutron-to-proton ratio 
$\sim 10$. 

Figure~3 shows examples of the disk structure around a Kerr black hole
with $a=0.95$ with accretion rate $0.2M_\odot$~s$^{-1}$, for three 
different values of viscosity parameter $\alpha$. The transitions at the
characteristic radii manifest themselves in the non-monotonic behavior
of the degeneracy parameter $\eta=\mu_e/kT$, $F^-/F^+$, $Y_e$, and $H/r$.
In particular the sharp maximum of $F^-/F^+$ appears at the ignition
radius. The disk is quite thin at $r<r_{ign}$ and cooled locally, 
$F^-\approx F^+$.

The high neutron richness has important implications for the global picture 
of GRB explosion.  When the neutron-rich material is ejected in a 
relativistic jet and develops a large Lorentz factor, the neutrons 
decay at distance $\sim 10^{16}$~cm and affect the observed explosion. 

Disks around rapidly rotating black holes produce a high neutrino flux.
It deposits energy above the disk (via neutrino 
annihilation $\nu+\bar{\nu}\rightarrow e^++e^-$) 
and can drive an outflow with power $L>10^{51}$~erg/s.
A detailed study of $\nu\bar{\nu}$ annihilation around a Kerr black hole is in 
preparation.

\begin{theacknowledgments}
This work was supported by NASA grant NAG5-13382.
\end{theacknowledgments}

{\begin{figure}[h] 
\centering 
$\begin{array}{cc} 
        \epsfig{file=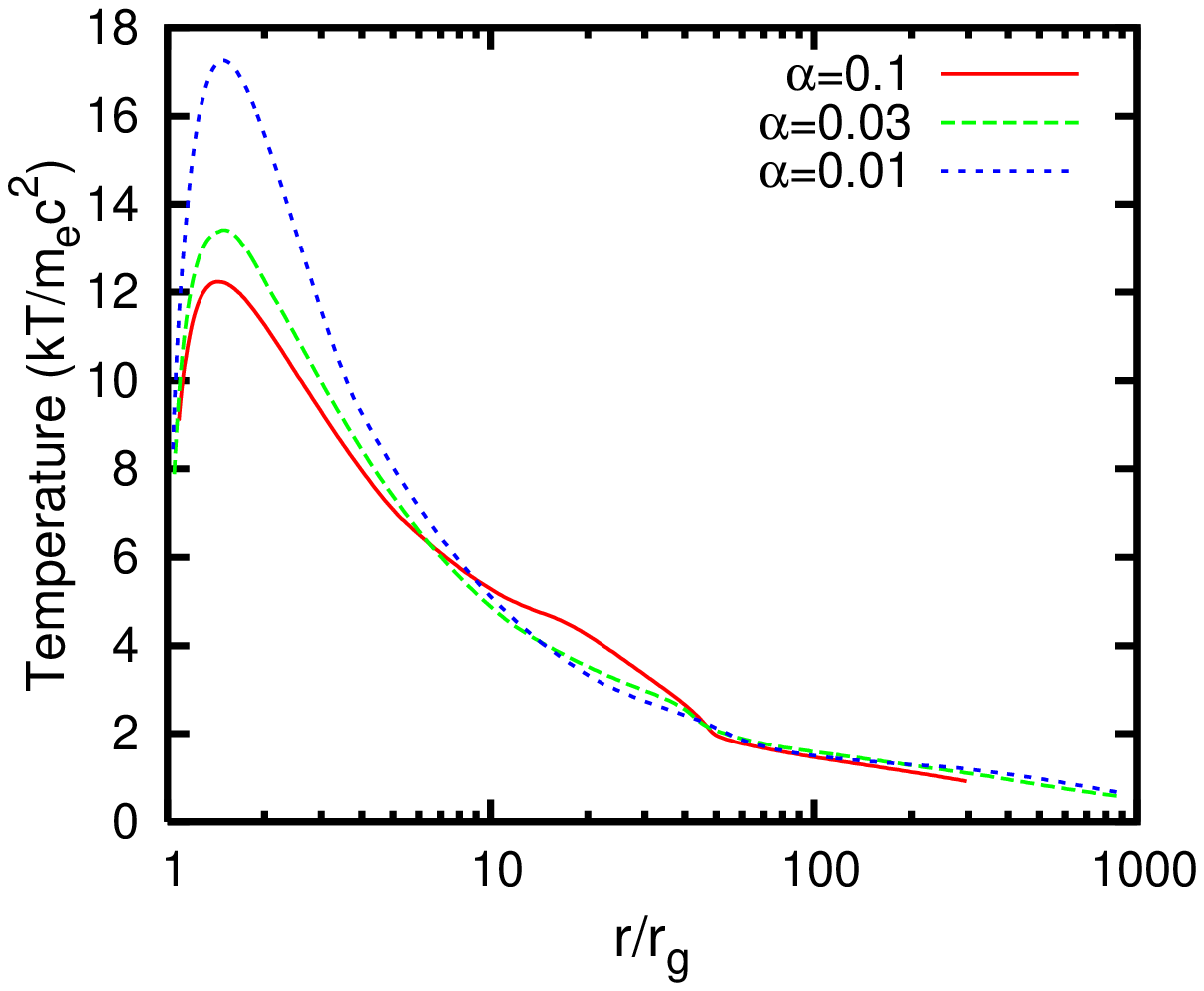, height=10.4pc, width=16pc} & 
        \epsfig{file=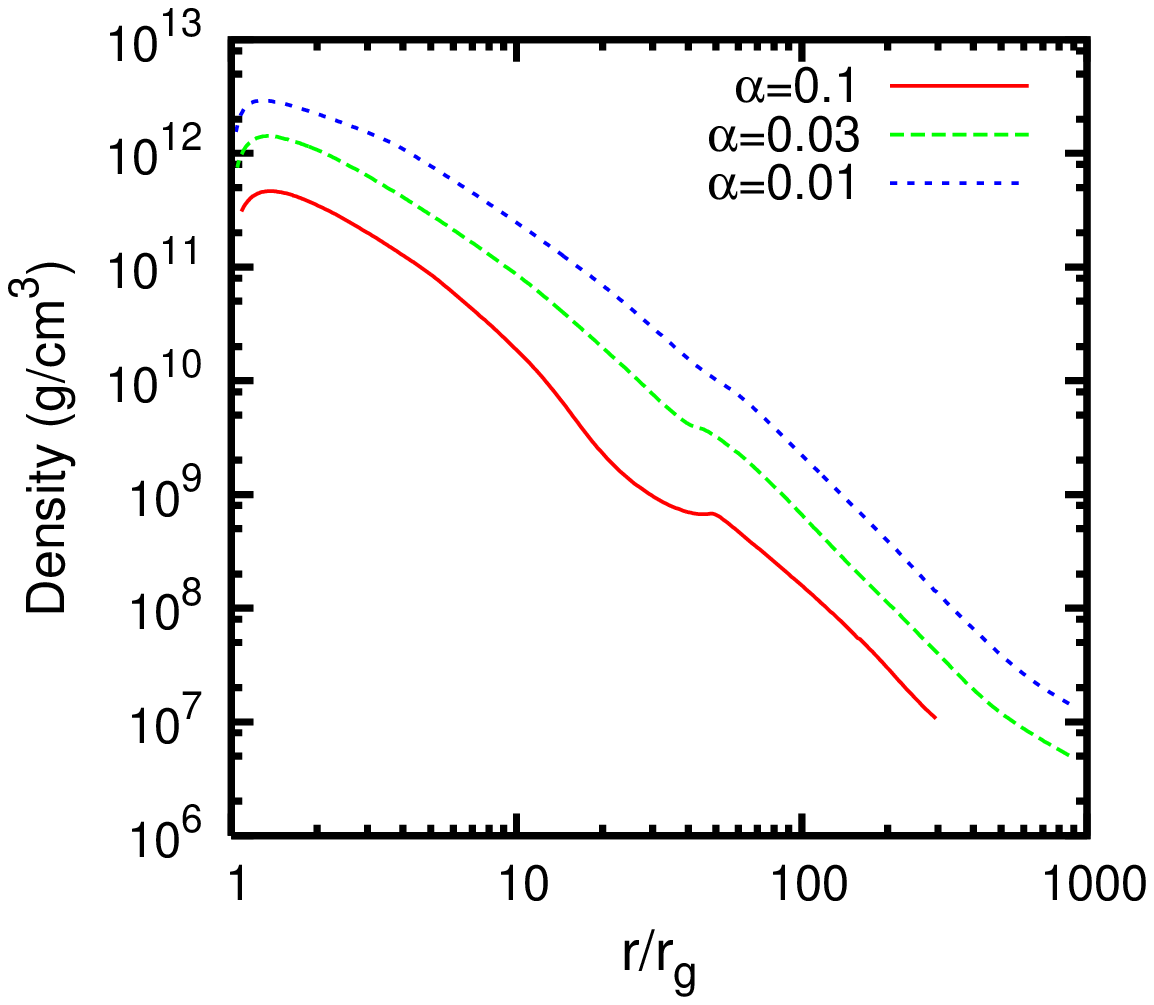, height=10.4pc, width=18pc} 
\end{array}$ 
\end{figure}
\begin{figure}[h] 
\centering 
$\begin{array}{cc} 
        \epsfig{file=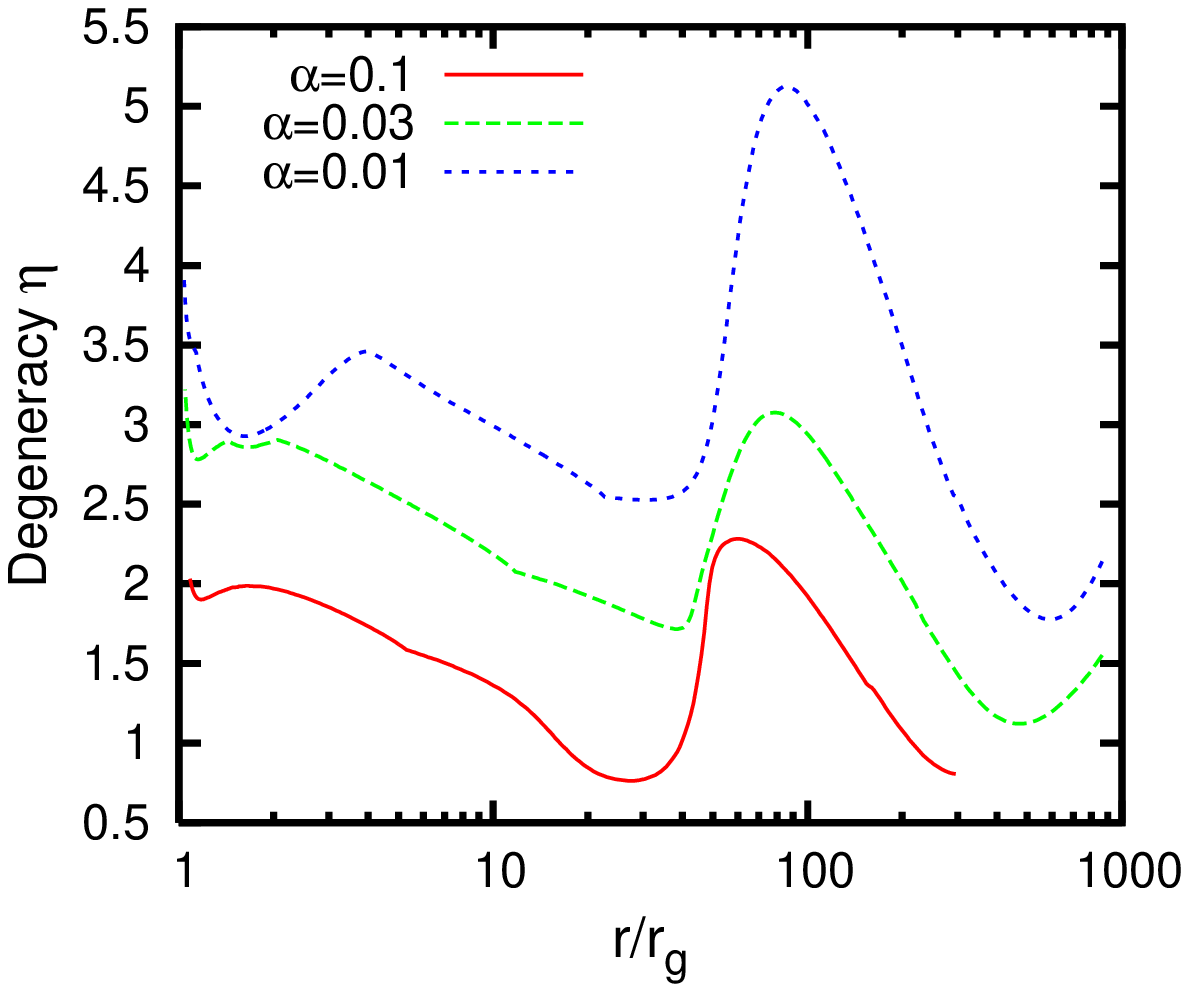, height=10.4pc, width=17pc} &
        \epsfig{file=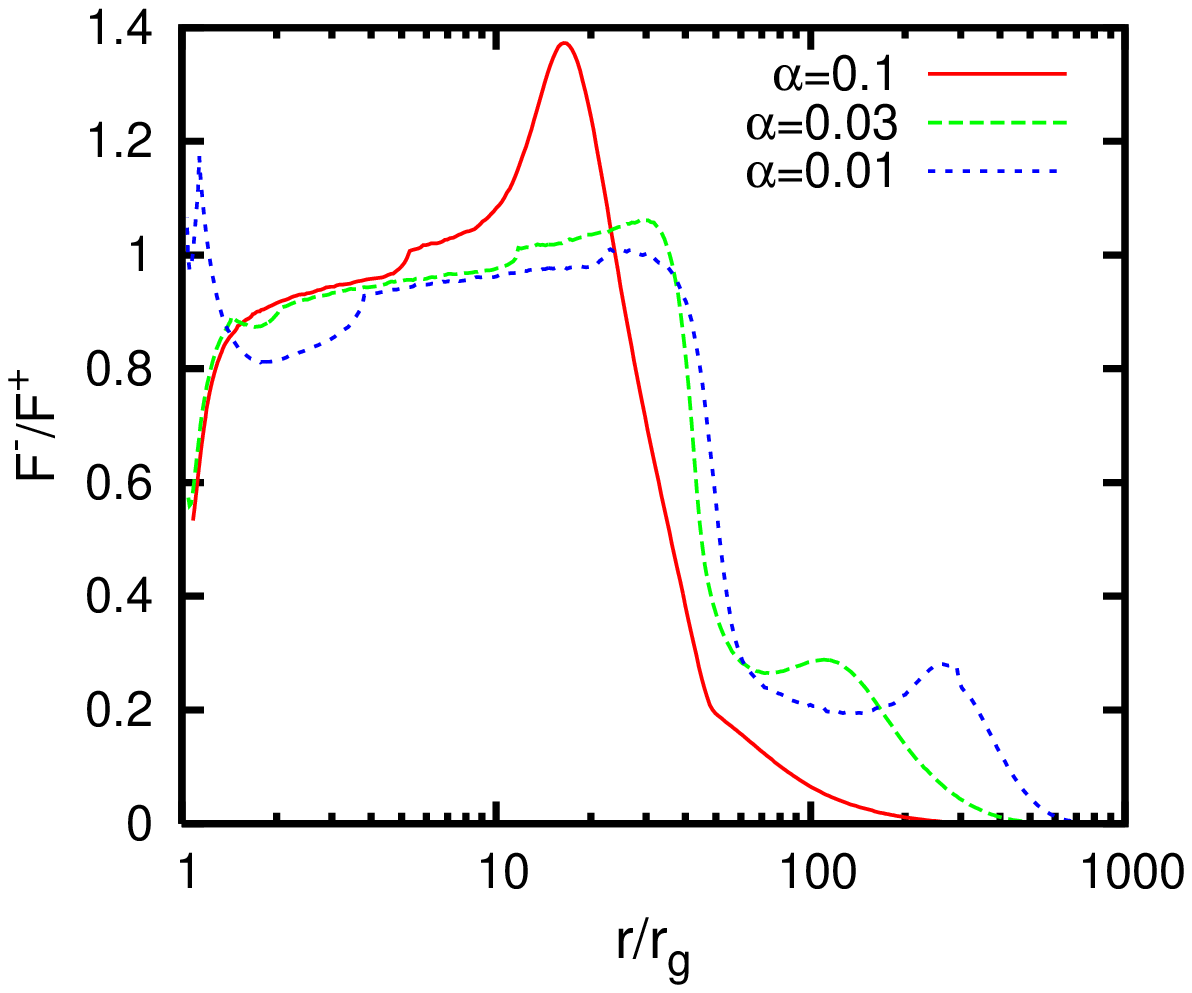, height=10.4pc, width=17pc} 
\end{array}$ 
\end{figure}
\begin{figure}[h] 
\centering 
$\begin{array}{cc} 
         \epsfig{file=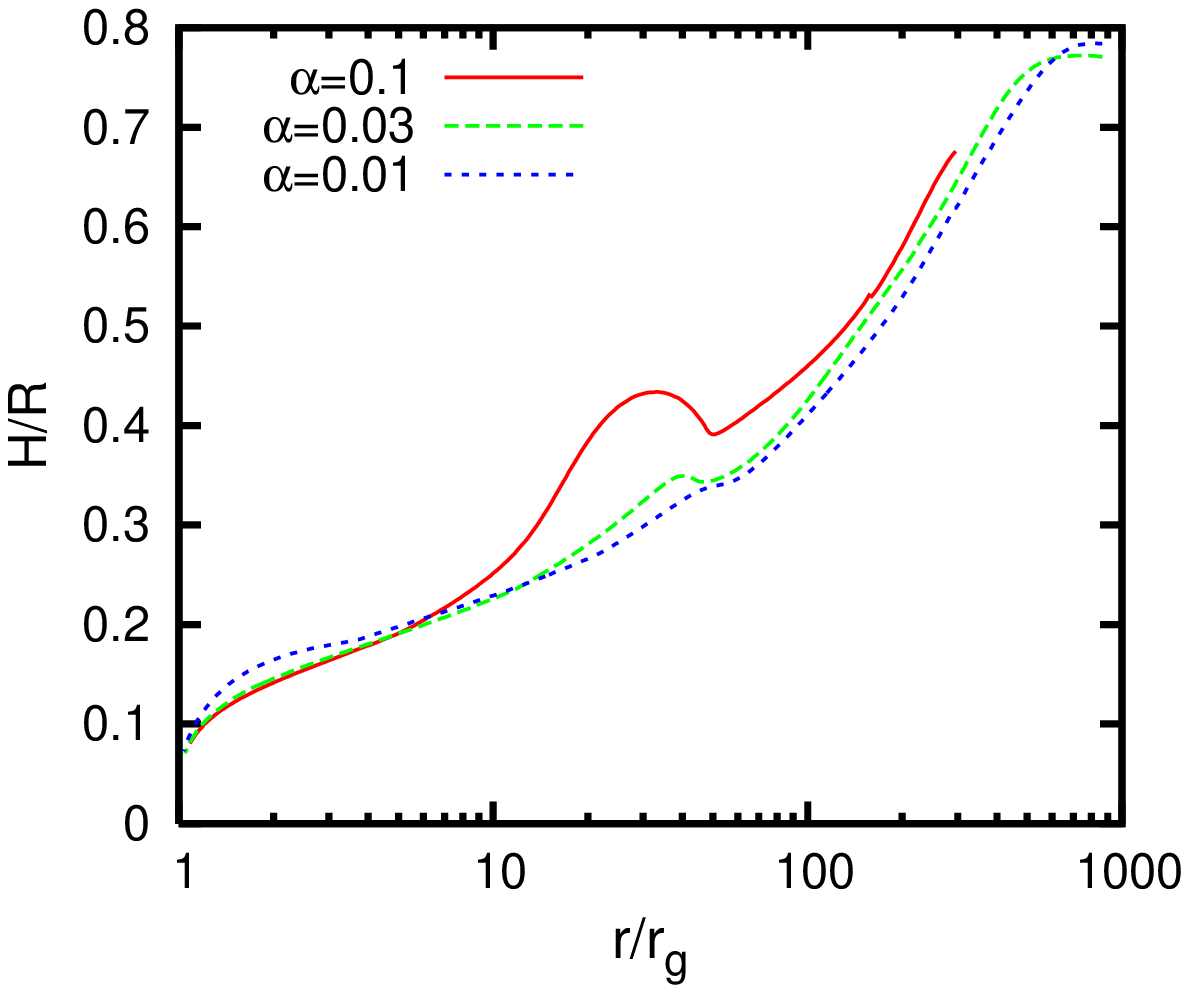, height=10.6pc, width=17pc} &
         \epsfig{file=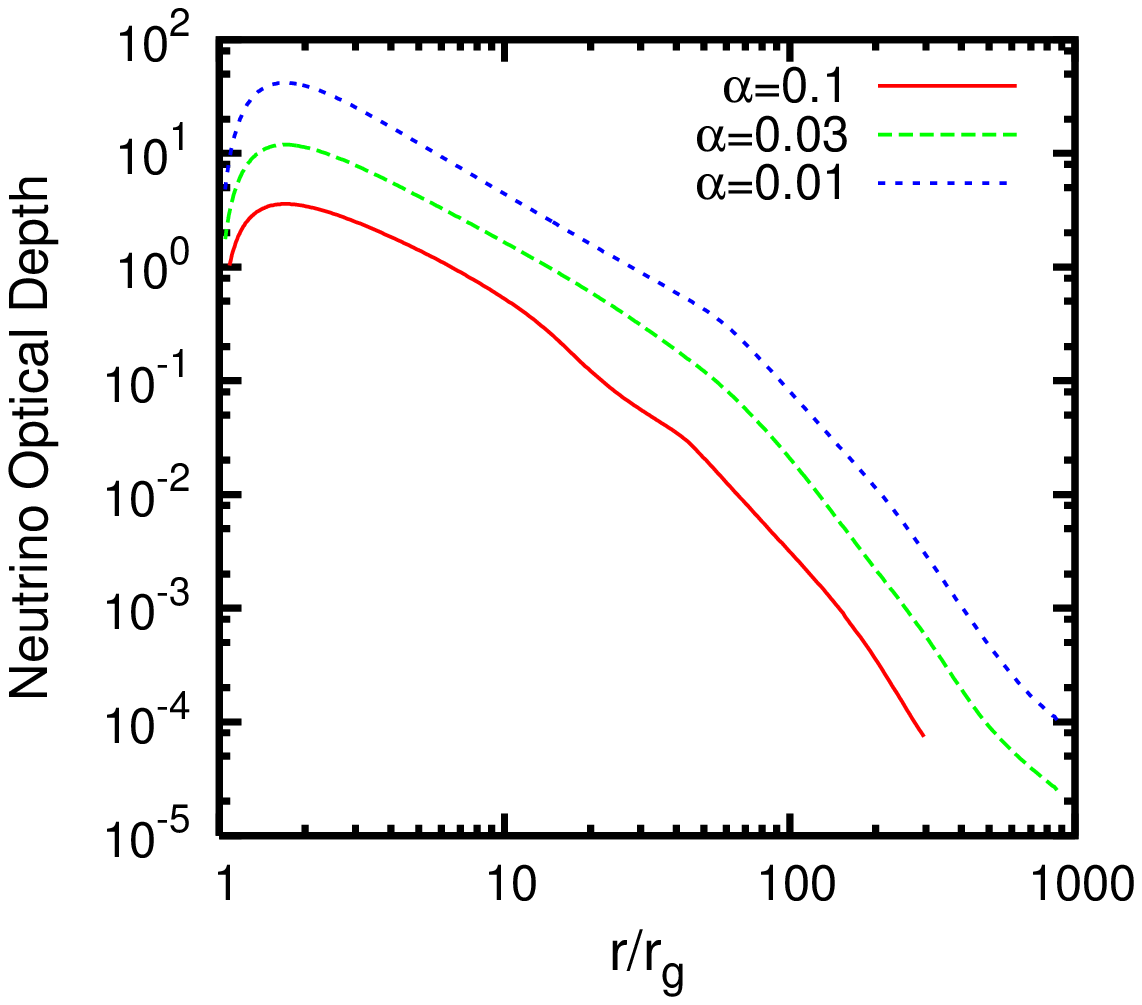, height=10.6pc, width=17pc} 
\end{array}$ 
\end{figure}
\begin{figure}[h] 
\centering 
$\begin{array}{cc} 
         \epsfig{file=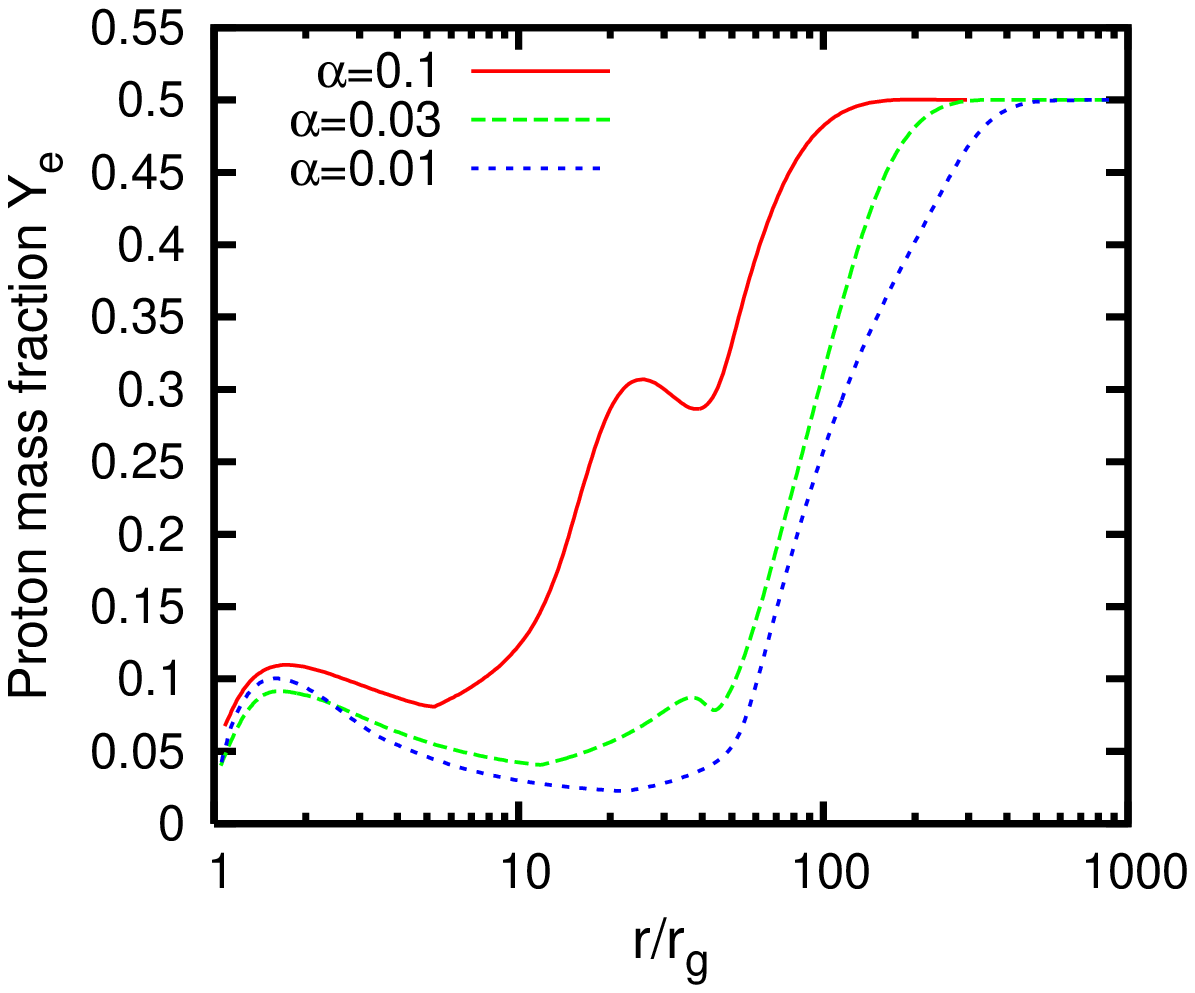, height=10.6pc, width=17pc} &
         \epsfig{file=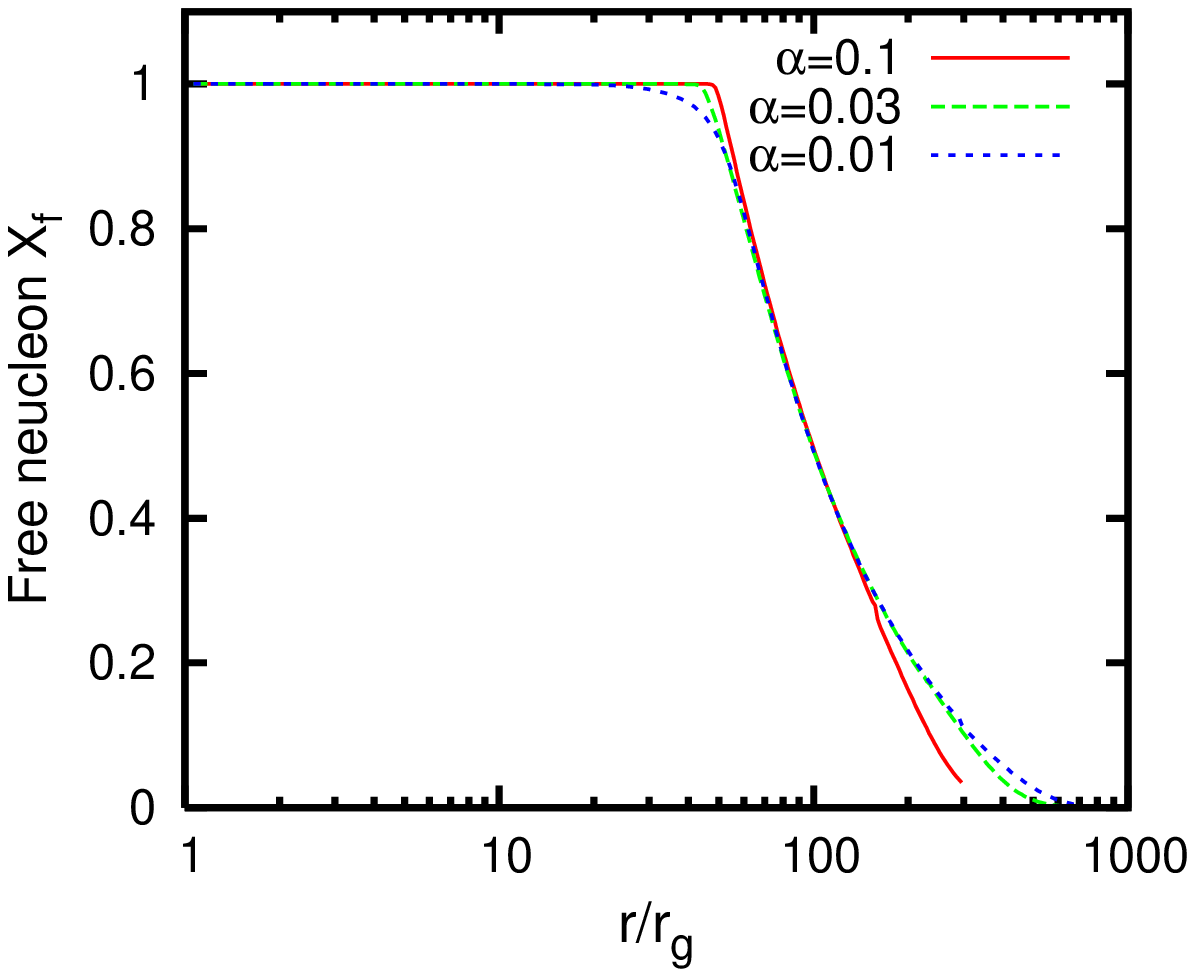, height=10.6pc, width=17pc} 
\end{array}$ 
\caption{Numerical solution for accretion disk with 
$\dot{M}=0.2 M_\odot s^{-1}$ 
around a black hole with mass $M=3M_\odot$ and spin parameter $a=0.95$. 
Three models are shown with viscosity parameter 
$\alpha=0.1$, $0.03$ and $0.01$. The degeneracy parameter $\eta$ is defined 
as $\mu_e/kT$, $X_{free}$ is the mass fraction of free nucleons. }
\end{figure} }


\begin{thebibliography}{9}

\bibitem{Popham} R. Popham, S.E. Woosley, \& C. Fryer, \emph{Astrophys.\ J.},
\textbf{518}, 356 (1999).

\bibitem{Belo} A.M. Beloborodov, \emph{Astrophys.\ J.},
\textbf{588}, 931 (2003).

\bibitem{Kohri} K. Kohri, R. Narayan, \& T. Piran, \emph{Astrophys.\ J.},
\textbf{629}, 341 (2005).

\end{thebibliography}
\end{document}